\documentclass[11pt]{article}
\usepackage{amssymb}

   \csname @addtoreset\endcsname{equation}{section}
    \csname @addtoreset\endcsname{figure}{section}
    \csname @addtoreset\endcsname{table}{section}

\newcounter{thanksnum}
\def\thanksnumber#1
{\setcounter{thanksnum}{\value{footnote}}\setcounter{footnote}{#1}%
                     \addtocounter{footnote}{-1}\footnotemark
                     \setcounter{footnote}{\value{thanksnum}}}
\def\newtheoremz#1{\@ifnextchar[{\@othmz{#1}}{\@nthmz{#1}}}

\def\@nthmz#1#2{%
\@ifnextchar[{\@xnthmz{#1}{#2}}{\@ynthmz{#1}{#2}}}

\def\@xnthmz#1#2[#3]{\expandafter\@ifdefinable\csname #1\endcsname
{\@definecounter{#1}\@addtoreset{#1}{#3}%
\expandafter\xdef\csname the#1\endcsname{\expandafter\noexpand
  \csname the#3\endcsname \@thmcountersepz \@thmcounterz{#1}}%
\global\@namedef{#1}{\@thmz{#1}{#2}}\global\@namedef{end#1}{\@endtheoremz}}}

\def\@ynthmz#1#2{\expandafter\@ifdefinable\csname #1\endcsname
{\@definecounter{#1}%
\expandafter\xdef\csname the#1\endcsname{\@thmcounterz{#1}}%
\global\@namedef{#1}{\@thm{#1}{#2}}\global\@namedef{end#1}{\@endtheoremz}}}

\def\@othmz#1[#2]#3{\expandafter\@ifdefinable\csname #1\endcsname
  {\global\@namedef{the#1}{\@nameuse{the#2}}%
\global\@namedef{#1}{\@thmz{#2}{#3}}%
\global\@namedef{end#1}{\@endtheoremz}}}

\def\@thmz#1#2{\refstepcounter
    {#1}\@ifnextchar[{\@ythmz{#1}{#2}}{\@xthmz{#1}{#2}}}

\def\@xthmz#1#2{\@begintheoremz{#2}{\csname the#1\endcsname}\ignorespaces}
\def\@ythmz#1#2[#3]{\@opargbegintheoremz{#2}{\csname
       the#1\endcsname}{#3}\ignorespaces}

\def\@thmcounterz#1{\noexpand\arabic{#1}}
\def\@thmcountersepz{.}
\def\@begintheoremz#1#2{ \trivlist \item[\hskip \labelsep{\bf #1\ #2}]}
\def\@opargbegintheoremz#1#2#3{ \trivlist
      \item[\hskip \labelsep{\bf #1\ #2\ (#3)}]}
\def\@endtheoremz{\endtrivlist}

\newtheorem{theorem}{Theorem}[section]
\newtheorem{lemma}{Lemma}[section]

\newtheorem{proposition}{Proposition}[section]
\newtheorem{corollary}{Corollary}[section]
\newtheorem{condition}{Condition}[section]
\newtheorem{definition}{Definition}[section]
\newtheorem{remark}{Remark}[section]

\def\M{{\bf M}}

\def\Fo{\F^{S,r}}

\def\e{\varepsilon}

\def\defi{\stackrel{{\scriptscriptstyle \Delta}}{=}}

\def\a{\alpha}
\def\d{\delta}

\def\O{\Omega}

\def\Y{{\cal Y}}
\def\F{{\cal F}}
\def\w{\widehat}

\def\Tr{{\rm tr\,}}

\def\R{{\bf R}}
\def\E{{\bf E}}
\def\P{{\bf P}}

\def\S{{\bf S}}
\def\Z{{\cal Z}}

\def\L{L}

\def\b{\beta}
\def\s{\delta}
\def\g{\gamma}

\def\W{{\cal W}}
\def\ww{\widetilde}
\def\X{{\cal X}}

\def\oo{\bar}
\def\s{\sigma}

\def\p{\partial}
\def\G{\Gamma}

\def\A{{\cal A}}
\def\M{{\cal M}}

\def\BB{{\rm B}}
\def\L{{\cal L}}

\newcommand{\be}{\begin{equation}}
\newcommand{\ee}{\end{equation}}
\newcommand{\bd}{\begin{displaymath}}
\newcommand{\ed}{\end{displaymath}}
\newcommand{\ba}{\begin{array}{ll}}
\newcommand{\ea}{\end{array}}
\newcommand{\baa}{\begin{eqnarray}}
\newcommand{\eaa}{\end{eqnarray}}
\newcommand{\baaa}{\begin{eqnarray*}}
\newcommand{\eaaa}{\end{eqnarray*}} \font\sm=cmr10

\def\y{\eta}
\def\YY{{\cal V}}
\date{January 12, 2005}
\title{Optimal solution of investment problems
via linear parabolic equations  generated by Kalman filter
\thanks{ Supported by NSERC of Canada under NCE Grant 30354 and
Research Grant 88051.}\footnote{{\it SIAM J. of Control and
Optimization} (2005) {\bf 44}, No. 4, pp. 1239-1258.}}
 \author{
Nikolai Dokuchaev\\
  {\sm   Department of Mathematics and
Statistics, University of Limerick, Ireland}}
\begin{document}\vspace{-0.5cm} \maketitle
\vspace{-1cm}\begin{abstract} We consider optimal investment
problems for a diffusion market model with non-observable random
drifts that evolve as an It\^o's process.  Admissible strategies
do not use direct observations of the market parameters, but
rather use historical stock prices. For a non-linear  problem with
a  general performance criterion, the optimal portfolio
strategy is expressed via the solution of a scalar minimization
problem and  a linear parabolic equation with coefficients
generated by the Kalman filter.
Abbreviated title: {\it Optimal solution of investment problems}
\end{abstract}
\section{Introduction}
The paper investigates an optimal investment problem  for a market
which consists of a locally risk free asset,  bond or bank account with
interest rate $r(t)$,  and a finite number,
$n$, of risky stocks.  We assume that the vector of stock
prices $S(t)$ evolves according to an   It\^o stochastic
differential   equation $dS_i(t)=S_i(t)[a_i(t)\ dt +
\sum_j \s_{ij}(t)\ dw_j(t)],$ $i=1,\ldots,n,$
with a vector of appreciation rates $a(t)$
and a volatility matrix $\s(t)$.
The problem goes
back to Merton (1969), who found  strategies which solve  the
optimization problem in which      $\E U(X(T))$ is to be
maximized, where $X(T)$ represents   the wealth at the final time
$T$ and where $U(\cdot)$ is a utility function.  If the market
parameters are observed, then the  optimal strategies (i.e.
current vector of stock holdings) are functions of the current
vector $(r(t),a(t),\s(t),S(t),X(t))$  (see, e.g., the survey in
Hakansson (1997) and Karatzas and Shreve (1998)). But in practice,
$a(t)$ and $\s(t))$ have to be estimated from historical stock
prices or some other observation process.     There are many
papers devoted to estimation of  $(a(\cdot),\s(\cdot))$, mainly based on
modifications of Kalman-Bucy filtering or the maximum likelihood
principle (see e.g. Lo (1988), Chen and Scott (1993), Pearson and
Sun (1994)). Unfortunately, the process $a(\cdot)$ is usually hard to
estimate in real-time markets,  because the drift term, $a(\cdot)$, is
usually overshadowed by the diffusion term, $\s(\cdot)$. On the other
hand, $\s(t)$ can, in principle, be found from stock prices. Thus,
there remains the problem of optimal investment with unobservable
$a(\cdot)$.
\par In fact, the problem is one of linear filtering. If $R_i(t)$
is the return on the $i$th stock, then $dR(t)= a(t) dt +\s(t)
dw(t),$ so the estimation of $a(t)$ given $\{R(\tau), \tau<t\}$
(or $\{S(\tau), \tau<t\}$) is a linear filtering problem. {If
$a(\cdot)$ is conditionally Gaussian,}  then the Kalman filter
provides the estimate which minimizes the  error in the mean
square sense.\par
A popular tool in optimal control and filtering  theory is the separation theorem.
This theorem has an analog in portfolio theory: it is the so-called {
``certainty equivalence principle'':} { agents who know the
solution of the optimal investment problem for the case of
directly observable $a(t)$ can solve the problem with unobservable
$a(t)$ by substituting $\E\{a(t)|S(\tau),\tau<t\}$} (see, e.g.,
Gennotte (1986)). Unfortunately, this principle does not hold in
the general case of non-log utilities (see  Kuwana (1995)). Note
that this principle is unrelated to the much more recent notion of
``certainty equivalent value" to be found in the work of Frittelli
(2000).
\par
Williams (1977), Detemple (1986),
Dothan and Feldman (1986), Gennotte (1986), Brennan (1998) solved
the investment problem using the Kalman-Bucy filter and dynamic
programming. By this method, the optimal strategy can be
calculated via solution of the
 Bellman parabolic equation; this equation is { non-linear}.
\par
{Karatzas} (1997), Karatzas and  Zhao (1998), {Dokuchaev and Zhou}
(2000), Dokuchaev and Teo (2000) have obtained optimal portfolio
strategies in general non-Gaussian setting, but only for case of
time independent coefficients.
\par
An approach  based on Malliavin calculus gives a possibility to
consider more general setting. Lakner (1995), (1998) assumes that
$S(\cdot)$ and $w(\cdot)$ have equal dimension (as we do),
 and that $r(\cdot)$ and $\s(\cdot)$
are deterministic. This again guarantees that the filtration of
$S(\cdot)$ is Brownian. Results from filtering theory give a
representation of the optimal portfolio, which is explicit in
terms of a conditional expectation of a Malliavin derivative when
the $a_i(\cdot)$ are Ornstein-Uhlenbeck processes independent of
$w(\cdot)$.
Karatzas and Xue (1990) assume that there are more Brownian
motions than stocks. They assume that $r(\cdot)$ and $\s(\cdot)$ are adapted to
the observable $S(\cdot)$. After projecting onto an $n$-dimensional
Brownian motion which generates the same filtration as $S(\cdot)$, they
obtain a reduced, completely observable model; existence of an
optimal portfolio follows, but the optimal strategy is, as usual,
defined only implicitly.
\par
We also consider the optimal investment problem with random and
unobservable $a(\cdot)$. Following Lakner (1998) and Rishel
(1999), we assume that $a(t)$ is a Gaussian process modelled by a
system of linear It\^o's equations. However, we consider a more
general case when $(a(\cdot),r(\cdot))$ may depend on the realized
returns (i.e., $b(\cdot)\neq 0$ in equation (\ref{aa}) below, and
$r(\cdot)$ is correlated with $S(\cdot)$).
 We  express the optimal strategy via solution of a
Cauchy problem (\ref{parab1}),(\ref{parab2}) for a {\it linear}
parabolic equation in $(n+1)$-dimensional vector space. Thus, we
propose  a simpler method than
 dynamic programming: the {\it nonlinear} parabolic Bellman
equation is replaced for  a  linear parabolic equation. Note that
the solution in Lakner (1998)  expresses the optimal strategy via
a conditional expectation of a random claim that depends on
$w(\cdot)$; the solution presented below is also based on the
martingale method but it is more constructive provided we can
solve the Cauchy problem (\ref{parab1}),(\ref{parab2}). Using the
technique of backward stochastic partial differential equations,
we prove existence and uniqueness of the solution for this Cauchy
problem. Furthermore, the most restrictive condition in Lakner
(1998) was  that the initial covariance of $a(0)$ is small enough
(the condition (\ref{lak}) below). We replace it by another
condition (\ref{quadr}) that depends on $U$: it is less
restrictive than (\ref{lak}) for some $U$'s and more restrictive
for others $U$'s. For some problems, our condition (\ref{quadr})
is automatically satisfied. In addition, we allow  correlated
$a(\cdot)$ and $w(\cdot)$.
\section{The Model and Definitions}\label{secD}
Consider a diffusion model   of a  market consisting of   a locally
 risk
free bank account or bond   with  price $B(t), $ ${t\ge 0}$, and
$n$ risky stocks with prices $S_i(t)$, ${t\ge 0}$,
$i=1,2,\ldots,n$, where $n<+\infty$ is given.   The prices of the
stocks evolve according to the following equations: \be \label{S}
dS_i(t)=S_{i}(t)\left(a_i(t)dt+\sum_{j=1}^n\s_{ij}(t)
dw_j(t)\right),  \quad t>0, \ee where   $w_i(t)$ are standard
independent Wiener processes, $a_i(t)$ are  appreciation  rates,
and $\s_{ij}(t)$ are volatility coefficients.   The initial price
$S_i(0)>0$ is a  given non-random constant.   The price of the
bond evolves according to   the following equation \be \label{B}
B(t)=B(0)\exp\left(\int_0^t r(t)dt\right), \ee where $B(0)$ is a
given constant which we take to be $1$ without loss of generality,
and $r(t)$ is the random process of the risk-free interest rate.
\par
We are given a  probability space
$(\O,\F,\P)$,
where $\O$ is a set of  elementary events, $\F$ is a
complete $\s$-algebra of events, and $\P$ is a probability measure.
\par
We introduce the vector  processes ($^\top$ denoted transpose)
$$
\begin{array}{ll}
w(t)=\left(w_1(t),\ldots,w_{n}(t)\right)^\top,\quad
S(t)=\left(S_1(t),\ldots,S_{n}(t)\right)^\top,\quad
a(t)=\left(a_1(t),\ldots,a_{n}(t)\right)^\top,
\end{array}
$$ and the matrix process $
\s(t)=\left\{\s_{ij}(t)\right\}_{i,j=1}^n. $ \par Let ${\bf
1}\defi(1,\ldots,1)^\top\in\R^n$, and $ \ww a(t)\defi
a(t)-r(t){\bf 1}.
$
\par
We define the  return to time $t$ by $dR_i(t)=dS_i(t)/S_i(t),\ R_i(0)=0$,  and introduce
the vector of returns
 $R(t)=\left(R_1(t),\ldots,R_{n}(t)\right)^\top$ and of excess returns
 $\ww R_i(t)=R_i(t)-\int_0^t r(s)\,ds.$
\par
Let $\{\Fo_t\}_{0\leq t\leq T}$ be  the filtration
 generated by the process
$(r(t),S(t))$ completed with the null sets of $\F$.
\par
Set $\ww S(t)\defi \exp\left(-\int_0^t r(s)ds\right)S(t)$.
\par We denote by $|x|$ the Euclidean norm of a vector $x\in\R^k$.
For an Euclidean space $E$, we  denote by $\BB([0,T];E)$  the set
of bounded measurable functions $f(t):[0,T]\to E$.  We denote by
$I_n$  the identity matrix in $\R^{n\times n}$. As usual, we say
that $A<B$ for symmetric matrices if the matrix $B-A$ is
definitely positive. We denote $\phi^-\defi\max(0,-\phi)$, and we
denote by $\mathbb{I}_{\{\cdot\}}$ the indicator function.
 \subsubsection*{The model for  $r,\s$, and $a$}
 \par To describe the
distribution of $\ww a(t)$, we shall use the model introduced in
Lakner (1998, p.84),  generalized for our case of  random $r$, non-constant
coefficients for the equation for $\ww a$, and
correlated $r$, $\ww a$, and $w$. We assume that we are given
measurable deterministic processes $\a(t)$, $\b(t)$, $b(t)$ and
$\d(t)$ such that \be\label{a-eq} d\ww a(t)=\a(t)[\d(t)-\ww
a(t)]dt+b(t)d\ww R(t)+\b(t) dW(t), \ee where $\a(t)\in\R^{n\times
n}$, $\b(t)\in\R^{n\times n}$, $b(t)\in\R^{n\times n}$,
$\d(t)\in\R^n$, and where $W$ is an $n$-dimensional Wiener process
in $(\O,\F,P)$. We assume that
$\a(t)$, $\b(t)$, $b(t)$, and $\d(t)$ are continuous in $t$ and such
that the matrix $\b(t)$ is invertible and $|\b(t)^{-1}|\le c$,
where $c>0$ is a constant. Further, we assume that $\ww a(0)$
follows an $n$-dimensional normal distribution with mean vector
$m_0$ and covariance matrix $\g_0$. The vector $m_0$ and the matrix
$\g_0$ are assumed to be known. We note that this setting covers
the case when $\ww a$ is an $n$-dimensional Ornstein-Uhlenbeck
process with mean-reverting drift.
\par Clearly, equation (\ref{a-eq}) can be rewritten as
\be\label{aa} d\ww a(t)=\Bigl(\a(t)\d(t)+[b(t)-\a(t)]\ww
a(t)\Bigr)dt+b(t)\s(t)dw(t)+\b(t) dW(t). \ee In addition, it can
be seen that $\ww R_i(t)$ evolve as \be \label{wR} d\ww R
_i(t)=\ww a_i(t)dt+\sum_{j=1}^n\s_{ij}(t) dw_j(t), \quad t>0. \ee
\par We
assume that the process
 $\s(t)$ is continuous in $t$, non-random and
such that    $ \s(t)\s(t)^\top\ge c_{\s}I_n$, where $c_{\s}>0$ is
a constant.
\par
Further, we assume that   $r(\cdot)=\phi_r(\ww R(\cdot),\Theta)$,
where $\Theta$ is a random element in a metric space $\X_r$, and
where $\phi_r:C([0,T];\R^n)\times \X_r\to \BB([0,T];\R)$ is a
measurable function, and $\Theta$ does not depend on
$(w(\cdot),W(\cdot),\ww a(0))$. In addition, we assume that the
process $r(t)$ is adapted to the filtration generated by $(\ww
R(t),\Theta)$. Note that  closed system (\ref{aa})-(\ref{wR}) for
the pair $(\ww a(t),\ww R(t))$ does not include $r(\cdot)$, and
$(\ww a(\cdot),\ww R(\cdot))$ does not depend on $\Theta$.
Therefore, the market model is well defined. The assumptions for
measurability of $r$ don't look very natural.  However, they cover
generic models when $r$ is independent on $\ww R$ or non-random,
and  we can still consider some models with correlated $r$ and
$\ww R$.
\par
Under these assumptions, the solution of (\ref{S}) is well
defined, but the market is incomplete. \par
Let  $\ww\phi_m(t,s)$, $m=0,1$, be the solution of the matrix equation
$$
\left\{\ba\frac{d\ww\phi_m}{dt}(t,s)=[m\cdot b(t)-\a(t)]\ww\phi_m(t,s),\\
\ww\phi_m(s,s)=I_n.
 \ea\right.
 $$
 Let \be\label{Km}
\ww K_{m}(t)\defi \int_0^t\ww\phi_m(t,s)b(s)\s(s)\s(s)^\top
b(s)^\top \ww\phi_m(t,s)^\top ds,\quad m=0,1.
\ee
\par We have that
$$ \ww a(t)=\ww\phi_1(t,0)\ww a(0)+ \int_0^t
\ww\phi_1(t,s)[\a(s)\d(s)ds+b(s)\s(s)dw(s)+\b(s)dW(s)]. $$  \par
It follows that $\ww K_{1}(t)$ is the covariance matrix for $\ww
a(t)$ calculated with $\b(t)\equiv 0$ and $\ww a(0)=0$. By the
linearity of (\ref{aa}),
 it follows that $\ww K_{1}(t)$ is
the conditional covariance for $\ww a(t)$ given
$(W(\cdot)|_{[0,t]},\ww a(0))$ or $(W(\cdot)|_{[0,T]},\ww a(0))$.
\par
Note that $\ww K_{m}(t)$ can be found as solutions of linear
equations that one can easy derive from (\ref{aa}) and (\ref{dwa})
(see, e.g., Arnold (1973), Chapter 8).
\par
We assume that $b$ is "small". More precisely, we assume that
there exists $\e>0$ such that \be\label{cov1} T\ww K_{m}(t)+\e
I_n< \s(t)\s(t)^\top \quad \forall t\in[0,T],\quad m=0,1. \ee
\subsubsection*{The risk neutral probability measure}
 Set $Q(t)\defi
(\s(t)\s(t)^\top)^{-1}$, and set \be \label{Z*}
\Z\defi\exp\left(\int_0^T [\s(t)^{-1}\ww a(t)]^\top dw(t)
+\frac{1}{2}\int_0^T \ww a(t)^\top Q(t)\ww a(t)dt\right). \ee
\begin{proposition}\label{propcov}
\be\label{cov2} \E\Bigl\{\exp \frac{1}{2}\int_0^T\ww a(t)^\top
Q(t)\ww a(t)dt\,\Bigl|\,W(\cdot),\ww a(0)\Bigr\}<+\infty
\qquad\hbox{a.s}. \ee
\end{proposition}
\par
By this proposition, the Novikov's condition is satisfied
conditionally,         $\E\{\Z^{-1}\,|\,W(\cdot),\ww a(0)\}=1$,
then $\E\Z^{-1}=1$.
\par
Define the (equivalent) probability measure $\P_*$  by
$d\P_*/d\P=\Z^{-1}$. Let $\E_*$ be the corresponding expectation.
\subsubsection*{The wealth and strategies}
 Let $X_0>0$ be the initial wealth at
time $t=0$,  and let $X(t)$ be the wealth at time $t>0$,
$X(0)=X_0$.  We assume that \be \label{X}
X(t)=\pi_0(t)+\sum_{i=1}^n\pi_i(t), \ee where the pair $(\pi_0(t),
\pi(t))$ describes the portfolio at time $t$.   The process
$\pi_0(t)$ is the  investment in the bond, $\pi_i(t)$ is the
investment in the   $i$th stock,
$\pi(t)=\left(\pi_1(t),\ldots,\pi_{n}(t)\right)^\top$, $t\ge 0$.
\begin{definition}
The process $\ww X(t)\defi \exp\left(-\int_0^tr(s)ds\right) X(t)$ is called the normalized
(or discounted) wealth.
\end{definition}
\par
Let $\S(t)\defi{\rm diag\,} (S_{1}(t),\ldots,S_{n}(t))$ and $\ww
\S(t)\defi{\rm diag\,} (\ww S_{1}(t),\ldots,\ww S_{n}(t))$ be
diagonal matrices with the corresponding diagonal elements. The
portfolio is said to be self-financing, if \be \label{self1}
dX(t)=\pi(t)^\top\S(t)^{-1}
dS(t)+\pi_0(t)r(t)dt=\pi(t)^\top dR(t)+\pi_0(t)r(t)dt. \ee
It follows from (\ref{X}) that for such portfolios \be\label{XX}
\ba
dX(t)=r(t)X(t)\,dt+\pi(t)^\top\left(\ww
a(t)\,dt+\s(t)\,dw(t)\right),\\
d\ww X(t)=B(t)^{-1}\pi(t)^\top d\ww R(t),\ea \ee
 so $\pi$ alone suffices to
specify the portfolio; the process $\pi_0$ is uniquely defined by
$\pi$ via (\ref{X}),(\ref{XX}); $\pi$ it is called a
self-financing strategy.
\begin{definition}
\label{def1} Let $\oo\Sigma$
 be the class of all
 $\Fo_t$-predictable
processes $\pi(\cdot)$ such that 
\begin{itemize}
\item
$\int_0^{T}\left(|\pi(t)^\top\ww
a(t)|^2+|\pi(t)^\top\s(t)|^2\right)\,dt<\infty\ \hbox{ a.s.}$
\item
there exists a constant $q_\pi$ such that
$\P\left(\ww X(t)-X_0\geq q_\pi,\forall t\in[0,T]\right)=1.$
 \end{itemize}
\end{definition}
A process $\pi(\cdot)\in \oo\Sigma$   is said to be an {\em
admissible} strategy with corresponding wealth $X(\cdot)$.
\par
For an admissible  strategy $\pi(\cdot)$, $X(t,\pi(\cdot))$ denotes
the corresponding    total wealth, and   $\ww X(t,\pi(\cdot))$  the
corresponding    normalized    total wealth. It follows that $\ww  X(t,\pi(\cdot))$ is a
$\P_*$-supermartingale with $\E_*\ww  X(t,\pi(\cdot))\leq X_0$ and
$\E_* |\ww  X(t,\pi(\cdot))|\leq |X_0|+ 2|q_\pi|.$
\par
Note that by definition, admissible strategies from $\oo\Sigma$
use  observations
of $r(t)$ and $S(t)$ only.
For these strategies, the processes $X(t)$ and $\ww X(t)$
are $\Fo_t$-adapted.
\par
The following definition is standard.
\begin{definition}
\label{repl}
Let $\xi$ be a given random variable.
An admissible  strategy   $\pi(\cdot)$
is said to replicate the claim $\xi$  if
$
X(T,\pi(\cdot))=\xi \quad\hbox{ a.s.}
$
\end{definition}
\section{Problem statement and preliminary results}
Let $T>0$, let $\w D\subset \R$ be convex and bounded below, and let $X_0\in \w D$ be given.
Let $U(\cdot):\w D\to\R\cup\{-\infty\}$ be such that $U(X_0)>-\infty$.
\par
We may state our general problem as follows:
Find   an admissible
self-financing strategy $\pi(\cdot)$
which solves the following optimization problem:
\be
\label{c}
\mbox{Maximize}\quad
\E U(\ww X(T,\pi(\cdot)))\quad\hbox{over}\quad
\pi(\cdot)\in\ \oo\Sigma
\ee
\be
\label{s}
\mbox{subject to }
\left\{
\begin{array}{l}
\ww X(0,\pi(\cdot))=X_0,
\\
\ww X(T,\pi(\cdot))\in\w D \quad
\hbox{a.s.}
\end{array}
\right. \ee The condition $\ww X(T,\pi(\cdot))\in\w D$ may
represent a requirement for a minimal normalized terminal wealth
if $\w D=[k,+\infty)$, $k>0$. This condition may represent also a
requirement for the normalized terminal wealth in goal achieving
problems if $\w D=[k_0,k_1]$, $k_0<k_1$.
\par
We assume that $U$, $X_0$ and $\w D$ satisfy the following two
conditions.
\begin{condition}
\label{ass01} There exists a measurable set $\Lambda \subseteq [0,\infty)$,
and a measurable function
$F(\cdot,\cdot):\,(0,\infty)\!\times\Lambda
\to \w D$
such that for each $z>0$,
 $\w x=F(z,\lambda)$ is a solution of
the optimization problem
\be
\label{UU}
\mbox{{\rm Maximize}}\quad
zU(x)-\lambda x
\quad\mbox{{\rm over  }} x\in \w D.
\ee
\end{condition}
Note that the usual concavity hypotheses imply this condition, but
more general utility functions are also covered. For example, this condition is
satisfied for the goal achieving problem when $U(x)$ is a step
function (see e.g. Karatzas (1997), Dokuchaev and Zhou (2000)).
\par
Let $\oo\Z\defi\E\{\Z|\Fo_T\}$.  Since $(\ww R(\cdot),\ww
a(\cdot))$ does not depend on $\Theta$, we have   that $\Z$ does
not depend on $\Theta$, and  $\oo\Z=\E\{\Z|\ww R(\cdot)\}$.   Let
$F(\cdot)$ be as in Condition \ref{ass01}.
\begin{condition}
\label{assla} There exists $\w\lambda\in \Lambda$  such that $
\E_*|F(\oo\Z,\w\lambda)|<+\infty$ and $\E_*
F(\oo\Z,\w\lambda)=X_0.$
\end{condition}
\par
We solve our problem in two steps using
the martingale approach. First we show that $\E U(F(\oo
\Z,\w\lambda))$  is an
upper bound for the expected utility of normalized terminal wealth
for $\pi(\cdot)\in\oo\Sigma$. Then we find a portfolio
$\w\pi(\cdot)$ which replicates the claim $B(T)F(\oo
\Z,\w\lambda)$. This establishes the optimality of $\w\pi(\cdot)$.
\subsection*{The optimal claim}
The following theorem is a reformulation of Theorem 2.5 from
Lakner (1998) under slightly more general conditions that allow
discontinuous  functions $F$ and $U$ such as step functions.
\begin{theorem} \label{ThM1} (Dokuchaev and Haussmann (2000)). With $\w\lambda$ as in
Condition \ref{assla}, let $\w\xi\defi F(\oo\Z,\w\lambda).$ Then
\begin{itemize} \item[(i)] $\E\, U^-(\w\xi)<\infty$, $\w\xi\in \w
D$ a.s.;
 \item[(ii)] $\E
U(\w\xi)\ge \E U(\ww X(T,\pi(\cdot)))$, $\forall
\pi(\cdot)\in\oo\Sigma$; \item[(iii)] The claim $B(T)\w\xi$ is
attainable in $\oo\Sigma$, and there exists a  replicating
strategy in  $\oo\Sigma$. This strategy is optimal for problem
(\ref{c})-(\ref{s}).
\end{itemize}
\end{theorem}
This theorem uses duality approach for constrained
optimization  that goes back to Lagrange, and
$\w\lambda$ is the corresponding Lagrange multiplier.
\begin{remark}\label{rem1}{\rm
Theorem 2.5 from Lakner (1998)  was stated  under some additional
assumptions that can be formulated in our notations as
\\
(i)  $b(t)\equiv 0$, $r$ is non-random, $r$, $\s$, $\a$, $\b$, $\d$ are
constant,  and $\w D=(0,+\infty)$; \\
(ii) $U$ is strictly concave  and continuously
differentiable on $(0,+\infty)$, and $ \lim_{x\to +\infty}
U'(x)=0$;
\\
(iii) there exists a function $J(\cdot):\w D\to\R$ such that
$J(\lambda/x)\equiv F(x,\lambda)$;
\\ (iv) $\E_*J(\lambda/\oo\Z)<+\infty$ for any  $\lambda >0$.}
\end{remark}
\subsection*{Solution via conditional expectation}
Let
$$\w a(t)\defi \E\bigl\{\ww a (t)\,\bigl|\,\Fo_t\bigr\}.
$$
Set $\ww\a(t)\defi \a(t)-b(t)$ and  $m_0\defi \E\ww a(0)$.\par
Let $\g(t)\in\R^{n\times n}$
be the unique solution (in the class of symmetric nonnegative definite matrices) of
the deterministic Riccati's equation \be\label{gamma}
\left\{\ba\frac{d\g}{dt}(t)=-[b(t)\s(t)^\top+\g(t)]Q(t)[b(t)\s(t)^\top+\g(t)]^\top-\ww\a(t)\g(t)
-\g(t)\ww\a(t)^\top+\b(t)\b(t)^\top,\\
\g(0)=\g_0.
 \ea\right.\ee
Here $\g_0\defi\E[\ww a(0)-m_0][\ww a(0)-m_0]^\top$.
In fact, $\g(t)=\E\left\{[\ww a (t)-\w
a(t)][\ww a(t)-\w
 a(t)]^\top\bigl|\Fo_t\right\}$.
\par
Let $A(t)\defi -\ww\a(t)-\g(t)Q(t)$, and let $\phi(t)$ be the solution of the matrix equation
$$
\left\{\ba\frac{d\phi}{dt}(t)=A(t)\phi(t),\\
\phi(0)=I_n,
 \ea\right.$$
where $I_n$ is the unit matrix in $\R^{n\times n}$.
\par
The following theorem is a reformulation of Theorem 4.3 from
Lakner (1998). It gives the solution of the investment problem via
conditional expectation of future values of some processes with
known evolution.
\begin{theorem} (Lakner (1998)). Let conditions (i)-(iv) in Remark \ref{rem1}  holds, let $U(x)$ be
twice differentiable on $(0,+\infty)$,  and let \be\label{lak}\Tr
\g_0+T\|\b\|^2<K_1,\quad K_1=\frac{1}{360T \|\s^{-1}\|^2K_0},\quad
K_0=\max_{t\in[0,T]}\|e^{-\a t}\|^2,\ee where $\|\cdot\|$ denotes
the Frobenius matrix norm, i.e.,
$\|\s^{-1}\|^2=\Tr[\s^{-1}{\s^{-1}}^\top]$. Further, let \be
\label{lak2} J(x)<K(1+x^{-5}),\quad -J'(x)<K(1-x^{-2})\ee for some
$K>0$. Then the optimal strategy is
$$
\pi(t)^\top=H(t)\oo \Z(t)\E\biggl\{J'(\w\lambda \oo
\Z)\oo\Z^{-2}\biggl[-\g(t)
[\phi(t)^\top]^{-1}\int_t^T\phi(s)^\top[\s^\top]^{-1}d\w
w(s)-\w a(t)\biggr]\biggl|\,\Fo_t\biggr\},
$$
where $H(t)\defi \w\lambda e^{r(t-T)} Q$ and $\w w(t)\defi
w(t)-\int_0^t\s^{-1}\w a(s)ds$.
\end{theorem}
\par We propose below another solution such
that the optimal strategy is presented via solution of a linear
deterministic parabolic equation. We replace  conditions
(\ref{lak}) by condition (\ref{quadr})  that can be less
restrictive and is always satisfied if $\w D$ is bounded. In addition, we
dropped condition (\ref{lak2}) and  the condition  that $(r,a)$ and $w$
are independent:
we allow $b(\cdot)\neq 0$ and $r=\phi_r(\ww R(\cdot),\Theta)$.
\section{Main results: Solution via linear parabolic equation}
Let $y(t)=(y_1(t),\ldots,y_{n+1}(t))=(\w a(t),y_{n+1}(t))$ be a
process in $\R^{n+1}$, where \baaa  \w a(t)&=&\E\{\ww
a(t)|\Fo_t\},\\ y_{n+1}(t)&=&- \frac{1}{2}\int_0^t\w a(s)^\top
Q(s)\w a(s)ds+ \int_0^t\w a(s)^\top Q(s)\,d\ww R(s).\eaaa
\par
Let functions  $f(\cdot):\R^{n+1}\times[0,T]\to\R^{n+1}$ and
$g(\cdot):\R^{n+1}\times[0,T]\to\R^{(n+1)\times n}$ be such that
$$
f(x,t)\defi \left(\ba[A(t)-b(t)\s(t)^\top Q(t)]\w x +\a(t)\d(t)\\
\hphantom{xxxi}-\frac{1}{2}\w x^\top Q(t)\w x\ea\right), \qquad
g(x,t)\defi \left(\ba
\hphantom{xi}[b(t)\s(t)^\top+\g(t)]Q(t)\\
\hphantom{xixxxi} \w x^\top Q(t)\ea\right).
$$
Here $ A(t)$ and $\g(t)$ are matrices defined above,  $\g(t)$ is the solution of
(\ref{gamma}), and
$$
x=(x_1,\ldots,x_{n+1})^\top=\left(\ba\hphantom{xi}\w x\\
x_{n+1}\ea\right),\qquad\w x=\left(x_{1},\ldots,x_{n}\right)^\top.
$$
\par
By Theorem 10.3 from Liptser and Shiryaev (2000), p.396, the
equation for $\w a(t)$ is \be\label{dwa} \left\{\ba d\w
a(t)=[A(t)\w a(t)-b(t)\s(t)^\top Q(t)\w a(t)+\a(t)\d(t)]dt
+[b(t)\s(t)^\top +\g(t)]Q(t)d \ww R(t),\\
\w a(0)=m_0.\ea\right. \ee By (\ref{dwa})-(\ref{dZ}), it follows
that $y(\cdot)$ is the solution of the It\^o's equation \be
\label{yy}
 \left\{
\begin{array}{ll}
dy(t)=f(y(t),t)dt+g(y(t),t)\,d\ww R(t),\\
y(0)=y_0,\end{array}\right. \ee with $$y_0=\left(\ba m_0\\ 0\ea\right)\in\R^{n+1}, \qquad m_0= \E\ww a(0).
$$
The function $f(y,t)$ here does not satisfy Lipshitz condition
with respect to $y\in\R^{n+1}$. However,  the solution of this
equation is uniquely defined. (It is shown in the proof of Lemma
\ref{lemma2} below that the solution of (\ref{yy})  can be
presented as a part of the unique solution of some It\^o's
equation with coefficients that are affine with respect to the
state variable).
\begin{lemma}
\label{lemma2} Let a function
$\Phi(\cdot):\R^{n+1}\to\R$ be such that
\begin{itemize} \item[(i)]  $\E_* \Phi(y(T))=X_0$;
\item[(ii)] $\Phi(x)$ is continuously twice differentiable;
\item[(iii)]
$\E_* \Phi(y(T))^2< +\infty$.
\end{itemize}Then
  there exists a unique classical solution $V:\R^{n+1}\times[0,T]\to\R$ of the
 boundary value problem
\begin{eqnarray}\label{parab1}
 \frac{\p V}{\p t}(x,t)+\frac{\p V}{\p x}(x,t) f(x,t)+
\frac{1}{2}\Tr \left\{\frac{\p^2 V}{\p x^2}(x,t)\, g(x,t)\s(t)\s(t)^\top g(x,t)^\top\right\}=0,\\
 \label{parab1'}
V(x,T)=\Phi(x).\end{eqnarray}
Further, the
processes $\ww X(t,\pi (\cdot))\defi V(y(t),t)$ and
$\pi(t)^\top\defi B(t) \frac{\p V}{\p x}(y(t),t)g(y(t),t)$,
 are uniquely defined as elements of the
spaces $C([0,T],L_2(\O,\F,P_*))$ and $L_2([0,T],L_2(\O,\F,P_*))$
respectively, and
 there exists a constant $C>0$ such that \be\label{estim2}
\sup_{t\in[0,T]}\E_* |\ww X(t,\pi (\cdot))|^2+ \E_*\int_0^T
B(t)^{-2} |\pi(t)|^2dt\le C\E_*|\Phi(y(T))|^2\ee for all these
$\Phi$. Furthermore, the strategy $\pi(t)=(
\pi_1(t),\ldots,\pi_{n}(t))$ belongs to $\oo\Sigma$  and
replicates the claim $B(T)\Phi(y(T))$ given the initial wealth
$X_0$ with the normalized wealth $\ww X(t)=V(y(t),t)$.
\end{lemma}
\par
Note that estimate (\ref{estim2}) reminds the Krylov-Ficera
estimate (see Theorem 5.3.3 from Rozovskii (1980)) or its
modification from Dokuchaev (1995)).
\par
Further, we have that  \be\label{dZ}
 d\oo\Z(t)=\w
a(t)\oo\Z(t)d\ww R(t). \ee This formula (\ref{dZ}) was derived in
Theorem 3.1 from Lakner (1998)  for the case when $\s$ is constant
and $b=0$. The proof for a non-constant $\s(t)$ and $b\neq 0$  can
be found in Dokuchaev and Haussmann (2000) and in Chapter 9 from
Dokuchaev (2002). It follows that \be\label{ylog}
y_{n+1}(t)=\ln \oo \Z(t).\ee
\par Introduce the function
$e(\cdot):\R^{n+1}\to\R$  such that $e(y)=\exp[y_{n+1}]$ for
$y=(y_1,\ldots,y_{n+1})^\top$. Note that $\oo\Z=e(y(T))$.
\par
 Let $V=V(x,t,\lambda):\,\R^{n+1}\times[0,T]\times\Lambda\to \R$ be the solution of
partial differential equation (\ref{parab1}) with the condition
 \be
 \label{parab2}
 V(x,T,\lambda)=F(e(x),\lambda).
 \ee
 The following result now is immediate.
 \begin{theorem}\label{ThMarkov}
 Let $\w\lambda$ be such as in  Condition \ref{assla}.
 Assume that  the function $F(\cdot,\w\lambda):\R\to\R$ is such that
is such that conditions (i)-(ii) of Lemma \ref{lemma2} are
satisfied with  $\Phi(x)\defi F(e(x),\w\lambda)$, and
  \be\label{quadr}
\E_* F(\oo\Z,\w\lambda)^2<+\infty.
 \ee
 Then there exists an unique
classical solution $V$ of problem (\ref{parab1})-(\ref{parab2})
for $\lambda=\w\lambda$, and there exists an admissible
self-financing strategy $\pi(\cdot)\in \oo\Sigma$ which replicates
the claim $B(T)F(\oo\Z,\w\lambda)$.
  This strategy is an optimal solution of problem
(\ref{c})-(\ref{s}), and \be \label{Xopt} \w\pi(t)^\top
=B(t)\frac{\p V}{\p x}(y(t),t,\w\lambda)b(y(t),t),\quad \ww
X(t,\pi (\cdot))=V(y(t),t,\w\lambda),
 \ee
\end{theorem}\par
Note that it is possible  that condition (\ref{quadr}) is not
satisfied but  the optimal claim $F(\oo\Z,\w\lambda)$ is
still replicable in the class of strategies $\oo\Sigma$. For example, let
 $U(x)\equiv \log x$, $X_0=1$,  and $(0,+\infty)\subseteq\w D$, then $\Lambda =(0,\infty)$,
$F(z,\lambda)= z/\lambda$, $\w\lambda=1$, and the strategy is
$\pi(t)^\top = B(t)\w a(t)^\top \oo\Z(t)Q(t)$ is replicating (and
optimal) even in the case when (\ref{quadr}) is not satisfied.
\section{Special cases}
Note that conditions (\ref{lak}) were imposed in Lakner (1998) with the only purpose
to ensure that
\be \label{lak3}\E_*\oo\Z^5<+\infty,\qquad \E_*\oo\Z^{-4}<+\infty. \ee
\par
Our condition (\ref{quadr})  for examples (i)-(iii) listed below is satisfied if
$\E_*\oo\Z^\mu<+\infty$ for some $\mu\in\R$. For example (i),
 condition (\ref{quadr}) is less restrictive than (\ref{lak3}) if $l<5/2$ and more restrictive
if $l>5/2$. For example (ii),
 condition (\ref{quadr})  is less restrictive than (\ref{lak3}) if $l<2$ and more restrictive
if $l>2$. For example (iii),
 condition (\ref{quadr})  is always  less restrictive than (\ref{lak3}).
These
examples are from Dokuchaev and
Haussmann (2001): \par
(i) {\it Power utility}. Assume
 $\w D=[0,+\infty)$, $X_0>0$, $U(x)= d^{-1}x^{d}$,
where either $d\in(0,1)$ or  $d< 0$. Then $\Lambda
=(0,\infty)$, ${F}(z,\lambda)=(z/\lambda)^{l}$, and $\w\lambda=
X_0^{-1/l}(\E_*\oo\Z^l)^{1/l}$, where $l={1/(1-d)}$.
\par
(ii) Assume
 $\w D=[0,+\infty)$, $U(x)= -x^{d}+x$, where $d=1+1/l$,
 and $l>0$ is an integer, $X_0>d^{-l}$. Then
$\Lambda =[0,\infty)$, ${F}(z,\lambda)=(1+\lambda/z)^{l}d^{-l}$,
$\w\lambda$ is a root of a polynomial of degree $l$.
\par
(iii) {\it Mean-variance utility}. Assume
  $\w D=\R$,
$U(x)= -kx^2+cx$, where   $k\in \R$ and   $c\ge 0$, $X_0>0$, then
$F(z,\lambda )=(c-\lambda/z)/(2k)$.
\par
We present below some sufficient conditions that ensure
$\E_*\oo\Z^\mu<+\infty$ and, therefore, can be useful for verifying (\ref{Phi}) .
\par
Let $\ww K(t)$ be  the covariance for
$\ww a(t)$ under the probability measure $\P_*$, and let $\w K(t)$ be the covariance for $\w a(t)$
under $\P_*$.
\begin{lemma} \label{lemma002} If $\mu\in[0,1]$, then
$\E_*\oo\Z^\mu<+\infty$.  Let $\mu<0$ or $\mu>1$. Then $\E_*\oo\Z^\mu<+\infty$
if there
exist $\e>0$ and $p>1$ such that at least one of the following conditions holds:
\begin{itemize}
\item[(i)] $\kappa(p)\w K(t)< \s(t)\s(t)^\top - \e I_n$
for $t\in[0,T]$, where  $\kappa(p)\defi q T(\mu^2p-\mu)>0$ with
$q\defi p(p-1)^{-1}$.
\item[(ii)] $\kappa(p)\ww K(t)< \s(t)\s(t)^\top - \e I_n$
for $t\in[0,T]$.
\end{itemize}
\end{lemma}
\par
It follows from Proposition \ref{propSS} below
that $\ww K(t)$ and $\w K(t)$ are the covariances
of the processes defined by (\ref{aa})
and (\ref{dwa}) respectively with $\ww
R(\cdot)$ replaced by $\ww
R_*(\cdot)$. Thus, these covariances
can be found as solutions of linear
deterministic equations that one can easy derive from (\ref{aa})
and (\ref{dwa})  (see, e.g., Arnold (1973), Chapter 8).
\section{Case of discontinuous $F$}
To proceed further, we shall need a special weighted $L_2$-space
with a weight defined via some parabolic equation. First, we
introduce the operator $$ \M(t) p\defi -\sum_{i=1}^{n+1}\frac{\p
}{\p x_i}\left(p(x) f_i(x,t)\right)+
\frac{1}{2}\sum_{i,j=1}^{n+1}\frac{\p^2 }{\p x_i\p
x_j}\left(p(x)\w g_{ij}(x,t)\right), $$ where $\w g\defi
g\s\s^\top g^\top$.
\par Let $\rho_i\in L_2(\R^{n+1})\cap C^2(\R^{n+1})$, $i=1,2$, be
given  such that $\rho_i(x)>0$ for all $x\in\R^{n+1}$ and
$\int_{\R^{n+1}}\rho_i(x)dx=1$.
\par
We consider the following parabolic equation \be\label{peq}
\left\{\ba \frac{\p p}{\p t}(x,t)=\M(t) p(x,t)+\rho_1(x), \quad
t\in[0,T],\\ p(x,0)=\rho_0(x). \ea \right. \ee This boundary value
problem has the unique classical solution $p(x,t)$ that is
continuous in $\R^{n+1}\times[0,T]$. Let $$\rho(x)\defi
\min_{t\in[0,T]}p(x,t).$$   We have that $$ p(\cdot,t)={\rm
G}(t,0)\rho_0+\int_0^t{\rm G}(t,s)\rho_1ds, $$ where ${\rm
G}(t,s)$ is the semigroup operator generated by (\ref{peq}) (with
$\rho_1\equiv 0$) and such that ${\rm G}(s,s)\rho_i\equiv\rho_i$.
We have that $({\rm G}(t,s)\rho_i)(x)>0$ for $t\in[s,s+\e)$ for
some $\e=\e(x,s)>0$. Hence $p(x,t)>0$ for all $x,t$, and
$\rho(x)>0$ for all $x\in\R^{n+1}.$ We shall use this $\rho$ as a
weight function.
\par
We have that $\rho \in L_2(\R^{n+1})\cap L_1(\R^{n+1})$, since
$|\rho(x)|\le|\rho_0(x)|$.
\par We  introduce the weighted space $L_{2,\rho}(\R^{n+1})$ with the
norm
$$\|u\|_{L_{2,\rho}(\R^{n+1})}\defi\left(\int_{\R^{n+1}}\rho(x)|u(x)|^2dx\right)^{1/2}.
$$ We introduce the space $\Y_k$ of functions
$u=\{u_i(x,t)\}_{i=1}^k:\,\R^{n+1}\times[0,T] \to\R^k$ with the
norm $$ \|u\|_{\Y_k}\defi
\Bigl(\sum_{i=1}^k\int_0^T\left\|u_i(\cdot,t)\right\|_{L_{2,\rho}(\R^{n+1})}^2dt\Bigr)^{1/2}.
$$ Further, we introduce the space $\W^1$ of functions
$u=u(x,t):\,\R^{n+1}\times[0,T] \to\R$ with the norm $$
\|u\|_{\W^1}\defi \left\|u\right\|_{\Y_1} + \left\|\frac{\p u}{\p
x}\, g\right\|_{\Y_n}.
$$
\par
Finally, we introduce the space $\W_{C}^1$ consisting of all
functions $u(\cdot)\in\W^1$ such that $u(\cdot)\in
C([0,T];L_{2,\rho}(\R^{n+1}))$ with the norm $$
\|u\|_{\W_{C}^1}\defi
\sup_{t\in[0,T]}\left\|u(\cdot,t)\right\|_{L_{2,\rho}(\R^{n+1})} +
\|u\|_{\W^1}. $$ The above space is a Banach space, since  the
weighted space $L_{2,\rho}(\R^{n+1})$ is a Hilbert space.
\par
 In fact,
the spaces $\Y_k$,  $\W^1$, and  $\W_{C}^1$,  are the completions
in the corresponding norms
 of the set of
smooth functions  $u:\R^{n+1}\times[0,T]\to\R^k$ or
$u:\R^{n+1}\times[0,T]\to\R$ respectively that have finite
support.
\begin{theorem}
\label{prop01} Let $p$ be the solution of (\ref{peq}), and let
$\W_{C}^1$ be the corrsponding space defined via the weight
$\rho(x)=\min_{t\in[0,T]}p(x,t)$. Let
$\Phi(\cdot):\,\R^{n+1}\to\R$ be a measurable function such that
\be\label{Phi}\int_{\R^{n+1}}p(x,T)\Phi(x)^2dx<+\infty.\ee Then
boundary value problem (\ref{parab1})-(\ref{parab1'})  admits a
unique solution $V\in \W_C^1$. Moreover, there exists a constant
$C>0$ independent on $\Phi(\cdot)$ and such that  \be\label{estim}
\|V\|^2_{\W_C^1}\le C\int_{\R^{n+1}}p(x,T)\Phi(x)^2dx.\ee
\end{theorem}
\par
Note that condition (\ref{Phi}) allows discontinuous $\Phi$.
\begin{remark}{\rm
The  definition of $\W_C^1$ ensures that problem
(\ref{parab1})--(\ref{parab1'}) can be stated in $\W_C^1$.
The functions $V$ and $(\p V/\p x) g$ are
measurable and $L_{2,\rho}$-integrable.
The equality in (\ref{parab1'}) is
the equality for elements of the space $L_{2,\rho}(\R^{n+1})$, it
is meaningful  since $V(\cdot,t)$ is continuous in $t$ in
$L_{2,\rho}(\R^{n+1})$.  The
equality in (\ref{parab1}) is the equality for elements of the
dual space ${\W^1}^*$, since  all components of
 $\frac{\p^2 V}{\p  x^2}(x,t)\, g(x,t)\s(t)\s(t)^\top
g(x,t)^\top$ belong to ${\W^1}^*$. }
\end{remark}                         \par
It follows from the proof of Theorem \ref{prop01} below that
$\|p(\cdot,T)\|_{L_1(\R^{n+1})}=2$. Hence (\ref{Phi}) is satisfied
for any bounded $\Phi$. In addition, it can be shown that
$\|p(\cdot,T)\|_{L_2(\R^{n+1})}\le C\sum_{i=1,2}
\|\rho_i(\cdot)\|_{L_2(\R^{n+1})}$, where $C>0$ is a constant that
does not depend on $\rho_i$. Therefore, (\ref{Phi}) is satisfied
for any $\Phi\in L_4(\R^{n+1})$.   \par Theorem \ref{prop01} gives
the possibility to present the optimal investment strategy via
solution of (\ref{parab1})--(\ref{parab1'}) for the case of
discontinuous $F$. An  example is the goal-achieving problem, when
 $\w D=[0,\infty)$, $X_0\in(0,\alpha)$, and
$U(x)=0$ if $0\leq x<\alpha$, $U(x)=1$ if $x\geq\alpha$.
Then  $\Lambda =(0,\infty)$,
$F(z,\lambda)=\alpha$ if $0<\lambda\le z/\alpha$, $F(z,\lambda)=0$ if
$\lambda>z/\alpha$, and (\ref{Phi})  holds for $\Phi(x)=F(e(x),\lambda)$
$(\forall\lambda)$.
\section{Appendix:
Proofs} \label{sec8}
{\it Proof of Proposition \ref{propcov} }. By Jensen' inequality, it
follows that $$
\ba
 \E\Bigl\{\exp \frac{1}{2}\int_0^T\ww a(t)^\top
Q(t)\ww a(t)dt\,\Bigl|\,W(\cdot),\ww a(0)\Bigr\}=\E\Bigl\{\exp
\frac{1}{T}\int_0^T\frac{T}{2}\ww a(t)^\top Q(t)\ww
a(t)dt\,\Bigl|\,W(\cdot),\ww a(0)\Bigr\}
\\\hphantom{xxxxxxxxxxxxxxxx}\le\frac{1}{T}\int_0^T\E\Bigl\{\exp
\frac{T}{2}\ww a(t)^\top Q(t)\ww a(t)dt\,\Bigl|\,W(\cdot),\ww
a(0)\Bigr\}.
 \ea $$
We have for definitely positive matrices that if $A>B>0$ then
$B^{-1}>A^{-1}$. By condition (\ref{cov1}) with $m=1$, it follows
that \be\label{k1} \ba \ww K_{0}(t)^{-1}&>T[\s(t)\s(t)^\top-\e
I_n]^{-1} =TQ(t)[I_n-\e Q(t)]^{-1}\\&=
TQ(t)\left[I_n+\sum_{k=1}^{+\infty}\{\e Q(t)\}^k\right]> TQ(t)+T\e
Q(t)^2>TQ(t)+M, \ea\ee where $M=M(\e)>0$ is a definitely positive
constant matrix.
 Clearly,  we can take $\e>0$ small enough to ensure
convergency of the series in (\ref{k1}).
\par
To complete the proof, we shall use the following fact. Let $\xi$
be a Gaussian $n$-dimensional vector,  $K_\xi\defi
\E(\xi-\E\xi)(\xi-\E\xi)^\top>0$. It is known that the probability
density function for $\xi$ is $C\exp[-\frac{1}{2}(x-\E\xi)^\top
K_\xi^{-1}(x-\E\xi)]$, where $C>0$ is a constant. It follows that
$\E\exp(\frac{1}{2}\xi^\top P \xi)<+\infty$ for any matrix $P
\in\R^{n\times n}$ such that $0<P<K_\xi^{-1}$.
 Then the proof follows from (\ref{k1}).
$\Box$ \par We introduce the process $$ \ww
R_*(t)\defi\int_0^t\s(s)\,dw(s).$$
\par
Let $n$-dimensional vector random process $\ww a_*(t)$ be defined
as the solution of
$$
d\ww a_*(t)=\Bigl(\a(t)\d(t)-\a(t)\ww
a_*(t)\Bigr)dt+b(t)d\ww R_*(t)+\b(t) dW(t),\quad \ww a_*(0)=\ww
a(0).
$$
\par
Set \be \label{Z**} \Z_*\defi\exp\left(\int_0^T [\s(t)^{-1}\ww
a_*(t)]^\top dw(t) -\frac{1}{2}\int_0^T \ww a_*(t)^\top Q(t)\ww
a_*(t)dt\right). \ee
\begin{proposition} \label{psi1}
There exists a measurable function $\psi:C([0,T];{\bf R}^n)\times
\BB([0,T];{\bf R}^n)\to{\bf R}$ such that   $\Z_*=\psi(\ww R_*(\cdot),\ww
a_*(\cdot))$ and $\Z=\psi(\ww
R(\cdot),\ww a(\cdot))$.
\end{proposition}
\par
{\it Proof}. Clearly, $\psi$ is defined by  \be \label{lZ}
 \log\Z_*=
\int_0^T \ww a_*(t)^\top{ Q}(t)\left( d\ww R_*(t)-\frac{1}{2}\ww
a_*(t)dt\right).\ee
$\Box$\par
 Let
$r_*(\cdot)\defi \phi_r(\ww R_*(\cdot),\Theta)$ and
 $B_*(t)\defi B(0)\exp\left(\int_0^tr_*(s)ds\right)$ ($\phi_r$ is defined in Section
 \ref{secD}).
 Let
 \be\label{eZ} \oo \Z_* \defi
\E\{\Z_* | \ww R_*(\cdot),r_*(\cdot)\}. \ee
\par
Let ${\cal T}\defi C([0,T];\R^n)\times\R^n$.
Clearly, there exists a measurable mapping $\A: [0,T]\times
C([0,T];\R^n)\times {\cal T}\to C([0,T];\R^n)$ such that $\ww
a_*(t)=\A(t,\ww R_*(\cdot),W(\cdot),\ww a(0))$ and
$\ww
a(t)=\A(t,\ww R(\cdot),W(\cdot),\ww a(0))$.
\par
We have that  $\oo\Z_* =\E\{\Z_* | \ww R_*(\cdot)\}=\oo\psi(\ww R_*(\cdot))$ and
$$
\oo\Z_* =\E\{\psi([\ww R_*(\cdot),\ww
a_*(\cdot)] |\ww R_*(\cdot)\}
 =\E\{\psi[\ww R_*(\cdot),
\A\bigl(\cdot,\ww R_*(\cdot),W(\cdot),\ww a(0)\bigr) ] |\ww R_*(\cdot)\}.
$$
By Proposition~\ref{psi1}, it follows that
$$
\oo\Z =\E\{\psi[\ww R(\cdot),\ww
a(\cdot)] |\ww R(\cdot)\}
 =\E\{\psi[\ww R(\cdot),
\A\bigl(\cdot,\ww R(\cdot),W(\cdot),\ww a(0)\bigr) ] |\ww R(\cdot)\}.
$$
Hence there exists a
measurable mapping $\oo\psi(\cdot): C([0,T];\R^n)\to \R$ such that
\be\label{lZb}\oo\Z=\oo\psi(\ww R(\cdot)),\quad
\oo\Z_*=\oo\psi(\ww R_*(\cdot)).\ee
\begin{proposition}
\label{propSS} Let a function $\phi:C([0,T];{\bf R}^n)\times \BB([0,T];{\bf
R}^n)\times \BB([0,T];{\bf R})\to{\bf R}$ be  such that
$\E\phi^-(\ww R(\cdot),\ww a(\cdot),r(\cdot))<+\infty$. Further, let
a function $\w\phi:C([0,T];{\bf R}^n)\times \BB([0,T];{\bf R})\to{\bf R}$ be
such that $\E\w\phi^-(\ww R(\cdot),r(\cdot))<+\infty$ .
Then \begin{eqnarray} \label{phiphi2} \E\phi(\ww R(\cdot),\ww
a(\cdot),r(\cdot)) =\E\Z_*\phi(\ww R_*(\cdot),\ww
a_*(\cdot),r_*(\cdot)), \\ \label{phiphi3} \E\w\phi(\ww
R(\cdot),r(\cdot)) =\E\oo\Z_*\w\phi(\ww R_*(\cdot),r_*(\cdot)), \\ \label{phiphi4} \E_*\w\phi(\ww R(\cdot),r(\cdot))
=\E\w\phi(\ww R_*(\cdot),r_*(\cdot)). \end{eqnarray}
\end{proposition}
\par
{\it Proof}.  By assumption $(\Theta,W(\cdot),\ww a(0))$ is
independent of $w(\cdot)$.  To prove  (\ref{phiphi2}) it suffices
to prove \be \label{cond} \E\left\{\phi(\ww R(\cdot),\ww
a(\cdot),r(\cdot))\Bigl|
        \Theta,W(\cdot),\ww a(0)\right\}
 =\E\left\{\Z_*\phi(\ww R_*(\cdot),\ww a_*(\cdot),r_*(\cdot))\Bigl|
        \Theta,W(\cdot),\ww a(0)\right\}\quad\hbox{a.s.}
\ee Thus, for the next paragraph, without loss of generality, we
shall suppose that $(\Theta,W(\cdot),\ww a(0))$ is deterministic,
since for each value of $(\Theta,W(\cdot),\ww a(0))$ we can
construct $\ww R,\ww R_*, \ww a, \ww a_*$.
\par
 By the
linearity of (\ref{aa}),
 it follows that $\ww K_{0}(t)$ defined by (\ref{Km}) is
the conditional covariance for $\ww a_*(t)$ given
 $(W(\cdot),\ww a(0))$.
Similarly to the proof of Proposition \ref{propcov}, it can be
shown that (\ref{cov1}) with $m=0$ ensures that
 $\E\{\Z_*|\Theta,W(\cdot),\ww a(0)\}=1$ and $\E\Z_*=1$.
 We define the probability measure $\oo\P$ by
$d\oo\P/d\P=\Z_*$. (Each value of $(\Theta,W(\cdot),\ww a(0))$
generates its own $\oo\P)$. By Girsanov's Theorem, the process $$
\oo w(t)\defi w(t)-\int_0^t \s(s)^{-1}\ww a_*(s)ds $$ is a Wiener
process under $\oo\P$. From this we obtain $$
\begin{array}{l}
d\ww R(t)=\A(t,\ww R(\cdot),W(\cdot),\ww a(0))dt+\s(t)dw(t),
\\
d\ww R_*(t)=\A(t,\ww R_*(\cdot),W(\cdot),\ww a(0))dt+ \s(t)d\oo w(t).
\end{array}
$$
Then for each value of $(\Theta,W(\cdot),\ww a(0))$ the processes
$(\ww R(\cdot),\ww a(\cdot),r(\cdot))$
and $(\ww R_*(\cdot),\ww a_*(\cdot),r_*(\cdot))$ have the same distribution on the probability
spaces defined by $\P$ and $\oo\P$ respectively, and  (\ref{cond}),
hence (\ref{phiphi2}) follows.
\par
Further, (\ref{phiphi3}) follows by taking conditional expectation in (\ref{phiphi2}).
Finally, using Proposition~\ref{psi1} and (\ref{phiphi2}),
$$\ba
\E_*\w\phi(\ww R(\cdot),r(\cdot))& =\E \Z^{-1} \w\phi(\ww R(\cdot),r(\cdot)) =\E
\psi(\ww R(\cdot),\ww a(\cdot))^{-1}
 \w\phi(\ww R(\cdot),r(\cdot))\\
 &=\E \Z_*\psi(\ww R_*(\cdot),\ww a_*(\cdot))^{-1}
\w\phi(\ww R_*(\cdot),r_*(\cdot)) =
\E \w\phi(\ww R_*(\cdot),r_*(\cdot)). \qquad \Box
\ea
$$
\par
 We turn now to Theorem~\ref{ThM1}. Define $\w\xi_*\defi F(\oo\Z_*,\w\lambda)$.
 It follows from (\ref{lZb}) that if we
define $\ww\phi$ by $\w\xi=\ww\phi(\ww R(\cdot))$, then $\w\xi_*=
\ww \phi(\ww R_*(\cdot))$.
\par
{\it Proof of Theorem \ref{ThM1}}. Let us show that $\E
U^-(\w\xi)<\infty$ so that $\E U(\w\xi)$ is well defined. For
$k=1,2,\ldots$, we introduce the random events $$
\O_*^{(k)}\defi\bigl\{-k\le U(\w\xi_*)\le 0\bigr\}, \quad
\O^{(k)}\defi\bigl\{-k\le U(\w\xi)\le 0\bigr\}, $$ along with
their indicator functions, $\mathbb{I}_*^{(k)}$ and
$\mathbb{I}_{}^{(k)}$, respectively. The number $\w\xi_*$ provides
the unique  maximum of the function $\oo\Z_* U(\xi_*)-\w\lambda
\xi_*$ over $\w D$, and $ X_0\in \w D$. By
Proposition~\ref{propSS}, we have,  for all $k=1,2,\ldots$, $$
\begin{array}{rl}
\E \mathbb{I}_{}^{(k)}
U(\w\xi)-\E\mathbb{I}_*^{(k)}\w\lambda\w\xi_* =\E
\mathbb{I}_*^{(k)}\left(\oo\Z_* U(\w\xi_*)-\w\lambda\w\xi_*\right)
\ge \E\mathbb{I}_*^{(k)}\left(\oo\Z_* U(X_0)-\w\lambda X_0\right)
\\
= \E \mathbb{I}_{}^{(k)}U(X_0)-\w\lambda X_0\P(\O_*^{(k)}) \ge
-|U(X_0)|-|\w\lambda X_0|>-\infty.
\end{array}
$$
Furthermore, we have that $\E|\w\xi_*|=\E_*|\w\xi|<+\infty$. Hence $\E
U^-(\w\xi)<\infty$.
\par
Now observe that for any $\pi\in\oo\Sigma$ we can apply
(\ref{phiphi3}) and (\ref{phiphi4}) to $U(\ww X^\pi(T))$ (and use
(\ref{lZb})) to obtain
$$
\ba \E U(\ww X^\pi(T))=\E_*\{\oo\Z U(\ww X^\pi(T))\}
&\le\E_*\{\oo\Z U(\ww X^\pi(T))-\w\lambda \ww X^\pi(T)\}
        +\w\lambda X_0\\
        &\le \E_*\{\oo\Z U(\w\xi)-\w\lambda \w\xi\}
        +\w\lambda X_0=\E_* \oo\Z U(\w\xi)= \E U(\w\xi).
\ea
$$
Thus (ii) is satisfied.
\par
Let us show (iii). Since $\s$ is non-random, hence $w$-adapted,
then $\w\xi_*=\w\phi(w(\cdot))$, where $\w\phi(\cdot):\BB([0,T];\R^n)\to
\R$ is a measurable functions. By the martingale representation
theorem,
$$
\w\xi_*=\E\w\xi_*+\int_0^Tf(t,w(\cdot)|_{[0,t]})^\top dw(t),
$$
where
$f(t,\cdot):\BB([0,t];\R^n)\to \R^n$ is a measurable function such that
$
\int_0^T|f(t,w(\cdot)|_{[0,t]})|^2dt<+\infty\quad \hbox{a.s.}
$
There exists  a unique  measurable function
$f_0(t,\cdot):\BB([0,t];\R^n)\to \R^n$ such that
$f(t,w(\cdot)|_{[0,t]})\equiv f_0(t,\ww R_*(\cdot)|_{[0,t]})$.
Thus,
$$
\w\xi_*=\E\w\xi_*+\int_0^Tf_0(t,\ww R_*(\cdot)|_{[0,t]})^\top dw(t)
=
\E\w\xi_*+\int_0^Tf_0(t,\ww R_*(\cdot)|_{[0,t]})^\top \s(t)^{-1}
d\ww R_*(t). $$ Proposition~\ref{propSS} implies that $\E\w\xi_*=
\E_*\w\xi=X_0$, and $$ \w\xi=X_0 +\int_0^Tf_0(t,\ww
R(\cdot)|_{[0,t]})^\top \s(t)^{-1}d\ww R(t). $$ Hence the strategy
$\w\pi(t)^\top=B(t)f_0(t,\ww R(\cdot)|_{[0,t]})^\top \s(t)^{-1}$
replicates $B(T)\w\xi$. It belongs to $\oo\Sigma$; in particular,
since $w$ and $\ww R$ generate the same sigma-algebra and $\w D$
is convex, then $\ww X(t,\pi(\cdot))=\E\left\{\w\xi\,|\,\ww
R(\cdot)|_{[0,t]}\right\}\in\w D$, hence bounded below. This
completes the proof of Theorem \ref{ThM1}. $\Box$
\par
{\it Proof of Lemma \ref{lemma2}}.
 Let $\YY\defi \R^n\times\R\times\R^{\frac{n(n+1)}{2}}$. Clearly,
   $\YY$ is a $\ww n$-dimensional linear vector
 space, where $\ww n\defi n+1+n(n+1)/2$. Let
$\ww y(t)=(\ww y_1(t),\ww y_2(t),\ww y_3(t))$ be a process in $\YY$
such that  $$ \ww y(t)=(y(t),\ww y_3(t))=(\ww y_1(t),\ww
y_2(t),\ww y_3(t))=\left(\w a(t), \ln\oo\Z(t),\w a(t)\w
a(t)^\top\right).
$$
The last equality is satisfied by (\ref{ylog}). It can be seen that the equation for $\ww y(t)$ is linear: $$ \ba
d\ww y_1(t)=[\w A(t)\ww y_1(t)+v(t)]dt + E(t)\,d\ww R(t),\\
d\ww y_2(t)=-\frac{1}{2}{\rm Tr}\{Q(t)\ww y_3(t)\}\,dt+\ww y_1(t)^\top Q(t)\,d\ww R(t),\\
d\ww y_3(t)=[\w A(t)\ww y_3(t)+\ww y_3(t)^\top\w A(t)^\top+v(t)\ww y_2(t)^\top
+\ww y_2(t)v(t)^\top +\frac{1}{2}\{E(t)\s(t)\s(t)^\top
E(t)^\top\}]dt
\\\hspace{8cm} +
E(t)\,d\ww R(t)\ww y_2(t)^\top
+y_2(t)\,d\ww R(t)^\top E(t)^\top.
\ea
$$
Here $\w A(t)$, $v(t)$, $E(t)$ are known deterministic functions in
$\R^{n\times n}$, $\R^n$ and $\R^{n\times n}$ respectively. In particular,
$\w A(t)=A(t)-b(t)\s(t)^\top Q(t)$.
Thus,
the equation for $\w y(t)$ can be rewritten as
\be \label{wy}
 \left\{
\begin{array}{ll}
d\ww y(t)=\ww f(\ww y(t),t)dt+\sum_{i=1}^n\ww g_i(\ww y(t),t)\,d\ww R_i(t),\\
\ww y(0)=\ww y_0,\end{array}\right. \ee with $\ww y_0=\left(m_0,0, m_0
m_0^\top\right)
$, and with some functions
 $\ww f(\ww x,t):\YY\times[0,T]\to\YY$ and
 $\ww g_i(\ww x,t):\YY\times[0,T]\to \YY$, $i=1,\ldots,n$,
 that are
 affine in $\ww x\in\YY$ with continuous in
$t$ coefficients. In particular,  $ \p\ww f(\ww x,t)/\p\ww x$ and $ \p\ww g_i(\ww x,t)/\p\ww x$
depend only on $t$, and they are uniformly bounded.
Hence (\ref{wy}) has an unique solution. Therefore, equation
(\ref{yy}) has the unique solution $y(t)$.
\par Let $\ww
V(\ww x,t)\defi \E_*\Phi(\oo y^{\tilde x,s}(T))$, where the process
$\oo y^{\tilde x,s}(\cdot)$ takes values in $\R^{n+1}$ and is such that $\ww y^{\tilde
x,s}(\cdot)=(\oo y^{\tilde x,s}(\cdot),\ww y_3^{\tilde x,s}(\cdot))$ is the
solution of (\ref{wy}) given the initial condition $\ww y(s)=\ww
x\in\YY$.
         Then $\ww V(\ww x,t)$ is
the classical solution of the boundary value problem for the
corresponding backward Kolmogorov's equation \be\label{wparab}
\left\{\ba \frac{\p\ww V}{\p t}(\ww x,t)+\L(t)\ww V(\ww x,t)=0,\quad
t\in[0,T],\\
V(\ww x,T)=\Phi(\ww x_1,\ww x_2),\ea\right.\ee where $\ww x=(\ww
x_1,\ww x_2,\ww x_3)\in \R^n\times\R\times\R^{\frac{n(n+1)}{2}}$,
and where $\L(t)$ is the second order differential operator on
functions $v:\YY\to\R$ generated by the Markov process $\ww y(t)$.
\par
Let $y^{x,s}(\cdot)=(y_1^{x,s}(\cdot),\ldots,y_{n+1}^{x,s}(\cdot))
$ be the solution of (\ref{yy}), and let $V(x,t)\defi
\E_*\Phi(y^{x,t}(T))$. Clearly,
$$
\ww y^{\tilde x,s}(t)\equiv (y^{x,s}(t),\w y^{x,s}(t) \w
y^{x,s}(t)^\top)=\left(y_1^{x,s}(t),\ldots,y_n^{x,s}(t),y_{n+1}^{x,s}(t),\,\w
y^{x,s}(t) \w y^{x,s}(t)^\top\right),
$$
 if
$$
\ba              \ww x=(x,\ww x_3)=(\w x,x_{n+1},\w x\w
x^\top),\quad \w x=(x_1,\ldots,x_n),\quad
x=(x_1,\ldots,x_n,x_{n+1})=(\w x,x_{n+1})\in\R^{n+1},\\ \ww x_3=\w
x\w x^\top\in\R^{n(n+1)/2},\quad \w
y^{x,s}(\cdot)=(y_1^{x,s}(\cdot),\ldots, y_{n}^{x,s}(\cdot)). \ea
$$
In that case, $V(x,t)\equiv\ww V(x_1,\w x_2,\w x_2\w x_2^\top)$,
where $x=(\w x,x_{n+1})$, $\w x\in\R^n$. Therefore, $V(x,t)$  is
the classical solution of problem (\ref{parab1})-(\ref{parab1'}).
\par
 Let
$y_*(\cdot)$ denotes the solution of (\ref{yy}) with $\ww
R(\cdot)$ replaced by $\ww
R_*(\cdot)=\int_0^\cdot\s(t)\,dw(t)$.
\par Set $\ww
X_*(t)\defi V(y_*(t),t)$. From (\ref{parab1})  and
 It\^o's Lemma, it follows that
$$
\ww X_*(T)=\ww X_*(t)+\int_t^TB_*(s)^{-1}\pi_*(s)^\top d\ww R_*(s), $$ where
$
\pi_*(t)^\top\defi B_*(t)\frac{\p V}{\p x}(y_*(t),t) g (y_*(t),t).
$
It follows that $\ww X_*(0)=V(y_*(0),0)=\E V(y_*(T),T)=X_0$ and
\be \label{lem2-2} d\ww X_*(t)=B_*(t)^{-1}\pi_*(t)^\top d\ww R_*(t),\quad \ww
X_*(T)=\Phi(y_*(T)). \ee Then $\ww X_*(t)=\ww\psi(t,\ww R_*)$
for some measurable $\ww\psi$, and the result follows if we observe
that
 $\ww X(t)=\ww\psi (t,\ww R)$ replicates the claim as desired
 for $\pi(t)^\top\defi B(t)\frac{\p V}{\p x}(y(t),t) g (y(t),t)$.
\par
To continue, we require some a priori estimates. Let
$\zeta_*(t)\defi B_*(t)^{-1}\s(t)^\top\pi_*(t)$.
 \par
We consider the conditional probability space given
$(\Theta,W(\cdot),\ww a(0))$. With respect to the
conditional probability space, it follows from (\ref{lem2-2})  that
\be
\label{4-2} \left\{
\begin{array}{ll} d\ww X_*(t)=\zeta_*(t)^\top dw(t),
\\
 \ww X_*(T)=\Phi(y_*(T)).
\end{array}
\right.
 \ee By
Proposition 2.2 El Karoui {\it et al} (1997), the (unique)
 solution $(\zeta_*(t),\ww X_*(t))$ of linear stochastic backward
equation (\ref{4-2})
 is a process in $L_2([0,T],L_2(\O,\F,P))\times C([0,T],L_2(\O,\F,P))$, and
there exists a constant $c_0$, independent of $(\Phi(\cdot),
\Theta,W(\cdot),\ww a(0))$ and
 such that
 $$\ba
\sup_{t\in[0,T]}\E\left\{ |\ww X_*(t)|^2\bigl|\, \Theta,W(\cdot),\,\ww a(0) \right\}
&+ \E\biggl\{\int_0^T|\zeta_*(t)|^2dt
\biggl|\, \Theta,W(\cdot),\,\ww a(0) \biggr\}\\ &\le
c_0\E\left\{\Phi(y_*(T))^2\bigl|\, \Theta,W(\cdot),\,\ww a(0) \right\}\quad\hbox{a.s}. \ea
$$
 Hence
\be \label{est1-2} \sup_{t\in[0,T]}\E |\ww X_*(t)|^2+
\E\int_0^TB_*(t)^{-2}|\pi_*(t)|^2dt \le c_1\E \Phi(y_*(T))^2, \ee
where $c_1>0$ is a constant that does not  depend on $\Phi(\cdot)$. Then
(\ref{estim2}) follows.
 This completes the proof. $\Box$
\par {\it Proof of Theorem \ref{ThMarkov}.}  Clearly,  the equation for $y(t)$ is
$$\left\{ \ba d\w a(t)=[A(t)\w y(t)-b(t)\s(t)^\top
Q(t)+\a(t)\d(t)]dt + \g(t)Q(t)\,d\ww R(t),\\
dy_{n+1}(t)=\frac{1}{2} \w a(t)^\top Q(t)\w a(t)dt-\w a(t)^\top
Q(t)\,d\ww R(t). \ea \right.$$ As in the proof above, it can be
shown that $\ww X(t)=V(y(t),t,\w\lambda)$ is the solution of some
equation (\ref{lem2-2}), i.e. it is the normalized wealth. Then
the proof follows. $\Box$
\par Let
${\cal N}_2$ be the set of all Gaussian processes $\oo a(t):[0,T]\times \O\to\R^n$
which are progressively measurable with respect to the filtration generated by
$[a(0),w(t),W(t)]$ and  such that $\E\int_0^T|\oo a(t)|^2dt<+\infty$.
For
$\oo a(\cdot)\in {\cal N}_2$, let
$$
Z(t,\oo a(\cdot))\defi \exp\left[\int_0^t  \oo a(s)^\top Q(s) d\ww R(s)
-\frac{1}{2}\int_0^t \oo a(s)^\top Q(s)\oo a(s)ds\right].
$$
\begin{proposition}\label{ppp}
Let $\oo a(\cdot)\in {\cal N}_2$,
let $p\in(1,+\infty)$, and let $\mu\in\R$, $\mu<0$ or
$\mu>1$.   Let $\oo K(t)$ be the covariance matrix of
$\oo a(t)$ under $\P_*$, and let
 $\kappa(p)\defi qT(\mu^2p-\mu)$, where
$q\defi p(p-1)^{-1}$.    Let  $\kappa(p)\oo K(t)< \s(t)\s(t)^\top -\e I_n$,
 where $\e >0$ is a constant.
 Then $\E_*Z(t,\oo a(\cdot))^\mu <+\infty$.
 \end{proposition}
 \par
{\it Proof of Proposition  \ref{ppp}}.
If $\mu\in [0,1]$, then $\E_*Z(t,\oo a(\cdot))^\mu <+\infty$ ( see Lakner (1998), p.93).
Therefore, we can assume without loss of generality that $\mu<0$ or $\mu>1$.
Clearly,
$$
\ba Z(t,\oo a(\cdot))^\mu= \exp\left[\mu\int_0^t  \oo a(s)^\top Q(s) d\ww
R(s) -\frac{\mu}{2}\int_0^t \oo a(s)^\top Q(s)\oo
a(s)ds\right]=\zeta(t)\zeta_0(t),\ea $$ where $$ \zeta(t)\defi
 \exp\left[\mu\int_0^t  \oo a(s)^\top Q(s) d\ww R(s)
-\frac{\mu^2p}{2}\int_0^t \oo a(s)^\top Q(s)\oo a(s)ds\right], $$
and
$$
\zeta_0(t)\defi\exp\left[\frac{\mu^2p-\mu}{2}\int_0^t \oo a(s)^\top
Q(s)\oo a(s)ds\right].$$ By H\"older inequality, $\E_*\Z^\mu\le
\left[\E_*\zeta(T)^p\right]^{1/p}\left[\E_*\zeta_0(T)^q\right]^{1/q}$.
\par
Similarly to the proof of Lemma A.1 from Lakner (1998), we have
that $\E_*\zeta(T)^p<+\infty$ because $\zeta(t)^p$ is a positive
local martingale with respect to $\P_*$, thus by Fatou's lemma it
is a supermartingale.
\par  By Jensen's inequality,
 \be\ba
\E_*\zeta_0(T)^q=\E_*\exp\left[q\frac{\mu^2p-\mu}{2}\int_0^T \oo
a(s)^\top Q(s)\oo
a(s)ds\right]\\
=\E_*\exp\left[\frac{1}{2T}\kappa(p)\int_0^T
\oo a(s)^\top Q(s)\oo a(s)ds\right]
\le
\frac{1}{T}\int_0^T\E_*\exp\left[\frac{1}{2}\kappa(p) \oo a(s)^\top Q(s)\oo
a(s)\right]ds.\hphantom{}\ea \label{eta1} \ee
 Remind that $Q\defi(\s\s^\top)^{-1}$, and
$\kappa(p)>0$. Similarly to (\ref{k1}), we obtain \be\label{k2}
\ba \oo K(t)^{-1}&>\kappa(p)[\s(t)\s(t)^\top-\e I_n]^{-1}
=\kappa(p)Q(t)[I_n-\e Q(t)]^{-1}\\&=
\kappa(p)Q(t)\left[I_n+\sum_{k=1}^{+\infty}\{\e Q(t)\}^k\right]>
\kappa(p)Q(t)+\kappa(p)\e Q(t)^2> \kappa(p)Q(t)+M_1, \ea\ee where
$M_1=M_1(\e)>0$ is a definitely positive constant matrix. (We can
take $\e
>0$ small enough to ensure convergency.)  Similarly to the proof of Proposition \ref{propcov},
it follows from (\ref{eta1}), (\ref{k2}) that
$\E_*\zeta_0(T)^q<+\infty$ and $\E_*Z(t,\oo a(\cdot))^\mu
<+\infty$.
 $\Box$
\par
{\it Proof of Lemma \ref{lemma002}}. If $\mu\in [0,1]$, then $\E_*\oo\Z^\mu <+\infty$
( see Lakner (1998), p.93). Therefore, we can assume without loss of generality that
$\mu<0$ or $\mu>1$. Note that  $\w a(\cdot)\in {\cal N}_2$. By Proposition \ref{ppp},
if (i) is satisfied then $\E_*\oo\Z^\mu <+\infty$.
\par Further, let (ii) be
satisfied. Clearly, $\ww a(\cdot)\in {\cal N}_2$. By Proposition
\ref{ppp} again, $\E_*Z(T,\ww a(\cdot))^\mu<+\infty$. By (\ref{eZ}),
$\oo\Z_*=\E\{Z(T,\ww a_*(\cdot))|\ww R_*(\cdot),r_*(\cdot)\}$. Hence by Jensen's inequality $\E_*\oo\Z^\mu \le
\E_*Z(T,\ww a(\cdot))^\mu<+\infty$.
$\Box$
\par
{\it Proof of Theorem \ref{prop01}}. Let $\tau$ be a random
variable that takes values in $[0,T]$ and such that
$\P(\tau=0)=1/2$ and $\P(\tau\in(t_1,t_2])=(t_2-t_1)/(2T)$ for
$0<t_1<t_2\le T$. Let $\eta_i\in L_2(\O,\F,\P,\R^{n+2})$ be random
vectors
 such that they have the probability
density functions $\rho_i(x)$, $i=0,1$.   We assume  that
$\tau,\eta_0,\eta_1,w,\Theta,W(\cdot),\ww a(0)$ are mutually
independent.
\par
Let $\eta\defi
\eta_0\mathbb{I}_{\{\tau=0\}}+\eta_1\mathbb{I}_{\{\tau>0\}}$, and
let $\y_*(\cdot)$ be the solution of the It\^o's equation
 \be \label{yyy}
 \left\{
\begin{array}{ll}
d\y_*(t)=f(\y_*(t),t)dt+ g(\y_*(t),t)d\ww R_*(t),\quad t>\tau,\\
\y_*(\tau)=\eta.\end{array}\right. \ee  Equation (\ref{peq}) is
the forward Kolmogorov's equation for the case  when time of birth
is distributed as $\tau$, and the vector $\y_*(t)$ has the
conditional probability density function $p(x,t)/2$ in the sense
that $\P(\y_*(t)\in \G,\ t\ge\tau)=1/2\int_{\G}p(x,t)dx$ for any
domain $\G\subset\R^{n+1}$, where $p$ is the solution of
(\ref{peq}).
\par
Note that we need random $\tau$ with the selected probability
density on $(0,T]$ to generate the free term in parabolic equation
(\ref{peq}).
\par
Assume that $\Phi(\cdot)\in C^2(\R^{n+1})$ and it has finite support.
Let $V(x,t)\defi
\E_*\Phi(y^{x,t}(T))$, where $y^{x,s}(\cdot)$ is the solution of
(\ref{yy}). Then  $V(x,t)$ is the classical solution of problem
(\ref{parab1})-(\ref{parab1'}).  Set $\ww Y_*(t)\defi V(\y_*(t),t)$.
From (\ref{parab1})  and
 It\^o's Lemma, it follows that
$$
\ww Y_*(T)=\ww Y_*(t)+\int_t^TB_*(s)^{-1}\varrho_*(s)^\top d\ww R_*(s),
\quad \tau \le t\le T, $$ where $\varrho_*(t)^\top\defi B_*(t)\frac{\p
V}{\p x}(\y_*(t),t) g (\y_*(t),t). $ Hence \be
\label{2} d\ww Y_*(t)=B_*(t)^{-1}\varrho_*(t)^\top d\ww R_*(t),\quad \ww
Y_*(T)=\Phi(\y_*(T)). \ee
\par
To continue, we require some  estimates.
Let $\w\zeta_*(t)\defi B_*(t)^{-1}\s(t)^\top\varrho_*(t)$.
\par
 Consider the conditional probability space given
$(\tau,\eta,\Theta,W(\cdot),\ww a(0))$.
With respect to the
conditional probability space, it follows from (\ref{2})  that \be
\label{4} \left\{
\begin{array}{ll} d\ww Y_*(t)=\w\zeta_*(t)^\top dw(t),\\
 \ww Y_*(T)=\Phi(\y_*(T)).
\end{array}
\right.
 \ee By Proposition 2.2 El Karoui {\it et al} (1997)) again, the (unique)
 solution $(\w\zeta_*(t),\ww Y_*(t))$ of stochastic backward
equation (\ref{4})
 is a process in $L_2([\tau,T],L^2(\O,\F,P))\times C([\tau,T],L^2(\O,\F,P))$
given
$(\tau,\eta,\Theta,W(\cdot),\ww a(0))$, and
there exists a constant $C_0$ that is independent of
$(\Phi(\cdot),\tau,\eta,\Theta,W(\cdot),\ww a(0))$, and such that
 $$\ba
\sup_{t\in[0,T]}\E\mathbb{I}_{\{t\ge\tau\}}\left\{|\ww
Y_*(t)|^2\bigl|\, \tau,\eta,\Theta,W(\cdot),\ww a(0) \right\}+
\E\biggl\{\int_0^T\mathbb{I}_{\{t\ge\tau\}}|\w\zeta_*(t)|^2dt\,
\Bigl|\, \tau,\eta,\Theta,W(\cdot),\ww a(0)
\biggr\}\\=\sup_{t\in[\tau,T]}\E\left\{|\ww Y_*(t)|^2\,\bigl|\,
\tau,\eta,\Theta,W(\cdot),\ww a(0) \right\}+
\E\biggl\{\int_\tau^T|\w\zeta_*(t)|^2dt \Bigl|\,
\tau,\eta,\Theta,W(\cdot),\ww a(0) \biggr\}\\
\hphantom{xxxxxxxxxxxxxxxxxxxxxxxxxxxxxxx} \le
C_0\E\left\{\Phi(\y_*(T))^2\,\bigl|\,\tau,\eta,\Theta,W(\cdot),\ww
a(0) \right\}\quad\hbox{a.s}. \ea $$
 Hence  there exists a constant $c_0$, independent of
$\Phi(\cdot)$ and such that \be \label{est1}
\sup_{t\in[0,T]}\E\mathbb{I}_{\{t\ge\tau\}}|\ww Y_*(t)|^2+
\E\int_0^T\mathbb{I}_{\{t\ge\tau\}}B_*(t)^{-2}|\varrho_*(t)|^2dt\le
c_0\E\Phi(\y_*(T))^2. \ee
\par
Let $\Phi(\cdot)$ be a general measurable function satisfying the
conditions specified in the theorem. Then, there exists a
sequence
 $\{\Phi^{(i)}(\cdot)\}$, where $\Phi^{(i)}(\cdot)\in C^2(\R^{n+1})$
are such that they all have finite support and
\be\label{20}
\ba
\E|\Phi^{(i)}(\y_*(T))-\Phi(\y_*(T))|^2
=\int_{\R^{n+2}}p(x,T)|\Phi^{(i)}(x)-\Phi(x)|^2dx \to
0\quad\hbox{as}\quad i\to\infty.\ea \ee
 Let $\ww Y_*^{(i)}(\cdot)$, $\varrho_*^{(i)}(\cdot)$, and
 $V^{(i)}(\cdot)$
be the corresponding processes and functions. Let $$
\Psi_{i,j}\defi \sup_{t\in[0,T]}\E\mathbb{I}_{\{t\ge\tau\}}|\ww
Y_*^{(i)}(t)-\ww Y_*^{(j)}(t)|^2+ \E\int_0^T
\mathbb{I}_{\{t\ge\tau\}}B_*(t)^{-2}|\varrho_*^{(i)}(t)-\varrho_*^{(j)}(t)|^2dt.
$$ By (\ref{est1})-(\ref{20}) and the linearity of (\ref{4}), it
follows that $$ \Psi_{i,j}\le
c_0\E|\Phi^{(i)}(\y_*(T))-\Phi^{(j)}(\y_*(T))|^2\to 0
\quad\hbox{as}\quad i,j\to\infty. $$ We have that
$V^{(j)}\in\W_C^1$, since they are bounded together with their
partial derivatives with respect to $x_1,\ldots,x_{n+1}$. Remind
that $0<\rho(x)\le p(x,t)$ for all $x,t$.   Furthermore, we have
that \baaa
\Psi_{i,j}&=&\sup_{t\in[0,T]}\int_{\R^{n+1}}p(x,t)|V^{(i)}(x,t)-V^{(j)}(x,t)|^2dx
\\\hphantom{}
&+& \int_s^T dt \int_{\R^{n+1}}p(x,t)\left|\left[\frac{\p
V^{(i)}}{\p x}(x,t)-\frac{\p V^{(j)}}{\p
x}(x,t)\right]g(x,t)\right|^2dx. \eaaa Hence
$
\|V^{(i)}-V^{(j)}\|^2_{\W_C^1}
\le \Psi_{i,j}\to 0
\quad\hbox{as}\quad i,j\to\infty.
$
Therefore,
$V^{(i)}$ is a
 Cauchy sequence in  $\W_C^1$, and it has the limit $V$ in $\W_C^1$.
 This $V$ is the desired solution, and (\ref{estim})
 is satisfied.
 This completes the proof. $\Box$
 \par
Note that it follows from the proof above that the sequences
$\bigl\{\ww Y_*^{(i)}(\cdot)\bigr\}_{i=1}^{\infty}$ and
 $\bigl\{\varrho_*^{(i)}(\cdot)\bigr\}_{i=1}^{\infty}$ are
Cauchy sequences in the spaces
$C([\tau,T];L^2(\O,\F,\P\{\cdot\,|\,\tau\}))$ and
$L_2([\tau,T];L^2(\O,\F,\P\{\cdot\,|\,\tau\}))$ respectively.
Hence the corresponding limits $ \ww Y_*(\cdot)$, $
\varrho_*(\cdot)$ exist and belong to these spaces given $\tau$.
\par
This paper presents development of some results and ideas that
grew up from our collaboration with Ulrich Haussmann during the
author's stay at Pacific Institute for the Mathematical Sciences,
Vancouver (see, e.g., Dokuchaev and Haussmann (2000)). The author
wish to thank Prof. U. Haussmann for the support and useful
discussion. The author also wishes to thank the anonymous referees
for their insightful comments which greatly strengthened the
paper.
 \subsection*{References}
$\hphantom{xxx}$L. Arnold, Stochastic differential equations. Theory and applications.
Wiley-Inter-Science. New York, 1973.
\par
M.J. Brennan, {The role of learning in dynamic portfolio
decisions}, {\it European Finance Rev.} {\bf 1} (1998), 295-306.
\par
R.-R. Chen,  L. Scott, {Maximum likelihood estimation for
a multifactor  equilibrium model of the term structure of
interest rates}, {\it J. Fixed Income}  {\bf 4} (1993), 14-31.
 \par
J.B. Detemple, { Asset pricing in an economy with incomplete
information}, {\it J. Finance} (1986) {\bf 41}, 369-382.
\par
N.G. Dokuchaev,  Probability distributions  of  Ito's processes:
estimations for density functions and for conditional expectations
of integral functionals. {\it Theory of Probability and its
Applications} {\bf 39} (1995), N 4. 662-670.
\par N.G.
Dokuchaev, U. Haussmann, Optimal portfolio selection and
compression  in an incomplete market  {\it Quant. Finance} {\bf 1}
(2001), 336-345.
\par
N.G. Dokuchaev, U. Haussmann,  Adaptive portfolio selection based
on Historical Prices.
Working paper. Presented in Quantitative Risk Management in Finance,
Carnegie Mellon University,
Pittsburgh. July 31 - August 5, 2000.
\par
N.G. Dokuchaev, K.L. Teo, { Optimal   hedging strategy for a
portfolio  investment problem with  additional constraints}, {\it
Dynamics of Continuous, Discrete  and Impulsive Systems}, {\bf 7}
(2000), 385-404.
 \par
N.G. Dokuchaev, X.Y. Zhou, {Optimal investment strategies
with bounded  risks, general utilities, and goal achieving},
{\it J. Mathematical Economics}, {\bf 35} (2000), 289-309.
\par
N.G. Dokuchaev, {\it Dynamic portfolio strategies: quantitative
methods and empirical rules for incomplete information.}
 Kluwer Academic Publishers, Boston, January 2002.
\par
U. Dothan, D. Feldman, {Equilibrium interest rates and multiperiod
bonds  in a partially observable economy}, {\it J. Finance} {\bf
41} (1986), 369-382.\par N. El Karoui, S. Peng, and M.C. Quenez,
{Backward stochastic differential equations in finance.} {\em
Mathematical Finance} {\bf 7} (1997),  1--71.
\par
M. Frittelli, {Introduction to a theory of value coherent with the
no-arbitrage principle}, {\it Finance and Stochastics}, {\bf 3}
(2000), 275-298.
\par
G. Gennotte, {Optimal portfolio choice under incomplete
information},  {\it J. Finance} {\bf 41} (1986), 733-749.
\par
N. Hakansson, {\em Portfolio analysis.} W.W. Norton, New
York, 1997.
\par
I. Karatzas, { Adaptive control of a diffusion to a goal and a
parabolic Monge-Amp\`ere type equation}, {\it Asian  J.
Mathematics} {\bf 1} (1997), 324-341.
\par
I. Karatzas, S.E.  Shreve,  {\it Brownian motion and
stochastic calculus},  2nd  ed.,  Springer-Verlag,  New York,
1991.
\par
I. Karatzas, S.E.  Shreve,  {\em Methods of mathematical finance},
Springer-Verlag, New York, 1998.
\par
I. Karatzas,  X.-X. Xue, { A note on utility maximization under partial
observations}, {\em  Mathematical Finance}, {\bf 1} (1991), no. 2, 57-70.
\par
I. Karatzas, X. Zhao,
{ Bayesian adaptive portfolio optimization},
{\em Handbook of  Mathematical Finance},
Cambridge University Press, 632-670, 2001.
\par
Y. Kuwana,
{Certainty equivalence and logarithmic utilities in consumption/investment
problems}, {\it Mathematical Finance} {\bf 5} (1995), 297-310.
\par
P. Lakner, Utility maximization with partial information.
{\em Stoch. Processes  Appl.} {\bf 56} (1995), 247-273.
\par
P. Lakner, Optimal trading strategy for an investor: the
case of partial information. {\em Stoch. Processes Appl.},
{\bf 76} (1998), 77-97.
\par
R.S. Liptser, A.N. Shiryaev, {\em Statistics of random
processes. I. General theory}, Berlin, Heidelberg, New York:
Springer-Verlag (2nd ed), 2000.
\par
A.W. Lo, { Maximum likelihood estimation of generalized
It\^o processes with discretely sampled data}, {\it
Econometrics Theory} {\bf 4} (1988), 231-247.
\par
R. Merton, { Lifetime portfolio selection under uncertainty: the
continuous-time case.} {\it Rev. Econom. Statist.} {\bf 51}
(1969), 247-257.
\par
N.D. Pearson, T.-S. Sun, { Exploiting the conditional density in
estimating  the term structure: An application to the Cox,
Ingresoll, and Ross model}, {\it J. Finance} {\bf 49} (1994),
1279-1304.
\par
R.W. Rishel, Optimal portfolio management with partial
observations and power utility function. In {\it Stochastic
Analysis, Control, Optimization and Applications}. (W.M.
McEneaney, G. Yin, Q. Zhang, eds.), Birkhauser, 1999, 605-620.
\par B.L. Rozovskii,
{\it Stochastic evolution systems. Linear theory and applications
to non-linear filtering.} Kluwer Academic Publishers.
Dordrecht-Boston-London, 1990.
\par
J.T. Williams, { Capital asset prices with heterogeneous beliefs},
{\it J.  Financial  Economics} {\bf 5} (1977), 219-240.
\par
J. Yong and  X. Y. Zhou, {\em  Stochastic controls:
Hamiltonian systems and HJB equations.} Springer-Verlag, New
York. 1999.

\end{document}